\documentclass[twocolumn,caps]{revtex4}

\usepackage{graphicx,color}
\usepackage{latexsym}
\usepackage{amsmath,amssymb}        
\usepackage{mathrsfs}
\usepackage{comment}
\usepackage{soul}

\begin{document}


\title{Characterizing the velocity of a wandering black hole and properties of the surrounding medium using convolutional neural networks}


\author{J. A. Gonz\'alez and F. S. Guzm\'an}
\affiliation{Laboratorio de Inteligencia Artificial y Superc\'omputo,
	      Instituto de F\'{\i}sica y Matem\'{a}ticas, Universidad
              Michoacana de  San Nicol\'as de Hidalgo. Edificio C-3, Cd.
              Universitaria, 58040 Morelia, Michoac\'{a}n,
              M\'{e}xico.}




\begin{abstract}
We present a method for estimating the velocity of a wandering black hole and the equation of state for the gas around, based on a catalog of numerical simulations. The method uses machine learning methods based on convolutional neural networks applied to the classification of images resulting from numerical simulations.
Specifically we focus on the supersonic velocity regime and choose the direction of the black hole to be  parallel to its spin. We build a catalog of 900 simulations by numerically solving Euler's equations onto the fixed space-time background of a black hole, for two parameters: the adiabatic index $\Gamma$ with values in the range [1.1, 5/3], and the asymptotic relative velocity of the black hole with respect to the surroundings  $v_{\infty}$, with values within $[0.2, 0.8]c$. For each simulation we produce a 2D image of the gas density once the process of accretion has approached a stationary regime.
The results obtained show that the implemented Convolutional Neural Networks are capable to classify correctly the adiabatic index $87.78\%$ of the time within an uncertainty of $\pm 0.0284$ while the
prediction of the velocity is correct $96.67\%$ of the times within an uncertainty of $\pm 0.03c$. We expect that this combination of a massive number of numerical simulations and machine learning methods will help analyze more complicated scenarios related to future high resolution observations of black holes, like those from the Event Horizon Telescope.

\end{abstract}


\maketitle


Observations of horizon size scale images of supermassive black holes are expected to be possible soon with the Event Horizon Telescope, in particular observations of Sgr A$^{*}$ \cite{Doeleman} and M87 \cite{Chatzopoulos}.  Recent observations have also shown  potential black holes wandering across different scenarios in the interstellar medium, which are based on line emission offsets of sources with respect to the center of their host galaxies. A recent example is the radio-loud QSO 3C 186, which is potentially a candidate of a  black hole recoil \cite{QSO3C}.  Other candidates for  wandering black holes are NGC 3718 \cite{Markakis2015}, the quasar SDSS 0956+5128 \cite{Steinhardt2012} and SDSS 1133 \cite{Koss2014}. At a different mass scale there is also the case of the so called Bullet, a potential high velocity feature detected in the W44 supernova remnant \cite{Yamada2017}. We foresee that the resolution of wandering black holes will also increase within the next few years. 

Combining these observations, we see that the study of traveling black holes is important in two landscapes. The first one is related to the progenitor of the moving black hole as the result of the merger of two original smaller black holes, as already shown for the candidate QSO 3C 186 in \cite{Lousto2017}.  The second landscape relates to the effects produced by the black hole on the gas around. In this article we study the second problem by analyzing idealized but feasible scenarios of the effects produced by a fast black hole on the gas around it. We set the problem as the accretion of  wind and present a method based on deep learning, to study two properties of the system -the velocity of the black hole and the equation of state of the gas- which we classify based on a catalog of images of the gas density that are the result of numerical simulations. 

{\it Description of the accretion scenario.} The evolution of a black hole traveling over a gas that will eventually emit the radiation to be detected can be modeled as a Bondi-Hoyle-Littleton (BHL)  accretion process, which is characterized by a uniformly distributed wind with asymptotic velocity falling onto the accretor \cite{Bondi}. The properties of the wind, including the streamlines and matter distribution,  differ properties in terms of the mass and spin of the black hole, as well as of the gas equation of state and the relative velocity of the black hole with respect to the gas around it.

An illustrative regime is that of a supersonic relative velocity. In such a scenario  the process eventually approaches a stationary state in which there is a high density shock cone formed in the downstream zone of the black hole \cite{Rezzolla,LoraGuzman,GraciaGuzman}. The shock cone has  a well defined structure in the scale size of a few black hole horizon radius. This represents a neat scenario allowing a machine learning algorithm to classify some of the properties of the system in terms of a sample of images of the accretion process.

{\it  Preparation of the catalog.} We carry out a set of simulations of the accretion of a supersonic wind characterized by an asymptotic velocity and an equation of state.
For this we assume the black hole has a normalized mass, $M=1$. We also assume that the black hole has spin. The reason for this is that in a scenario in which the wandering black hole is the result of the merger of two black holes of the same mass and nearly symmetric spins, the black hole is kicked out with spin due to the orbital premerger angular momentum of the binary, and moreover, traveling in a direction nearly perpendicular to the orbital plane \cite{Gonzalez,Sperhake,Campanelli}. 
In this scenario the spin and the velocity of the black hole are nearly parallel. For this reason, in our idealized scenario we assume the 
black hole to have a spin of $\vec{S}=0.5\hat{z} M^2$ and relative velocity with respect to a gas of $\vec{v}_{BH}=v_{\infty}\hat{z}$.

For the simulation of the accretion we describe the black hole using  Kerr-Schild coordinates because the accretion process can be followed across the horizon and has been shown to be adequate to study the process \cite{LoraGuzman,GraciaGuzman}. Further, it allows the use of excision to remove a chunk of the domain inside the horizon of the hole in order to avoid the singularity during the evolution of the gas \cite{Nerozzi}.

The fluid is modeled using the set of relativistic Euler equations on a curved background for a perfect fluid of pressure $p$ and rest mass density $\rho$. We assume a gamma-law equation of state $p=(\Gamma -1)\rho\epsilon$, where $\epsilon$ is the internal energy of the fluid. For the numerical evolution we used the Cactus Einstein Toolkit \cite{ETK} in a test field mode, that is, we did not evolve the space-time geometry because the accretion of a very diluted wind is not expected to  significantly increase the mass of the black hole during the process. The numerical methods for the evolution are the standard high resolution shock capturing methods using the HLLE flux formula, implemented on a discretized numerical cubic domain \cite{whisky}.

\begin{figure}
\centering
\includegraphics[width= 5cm]{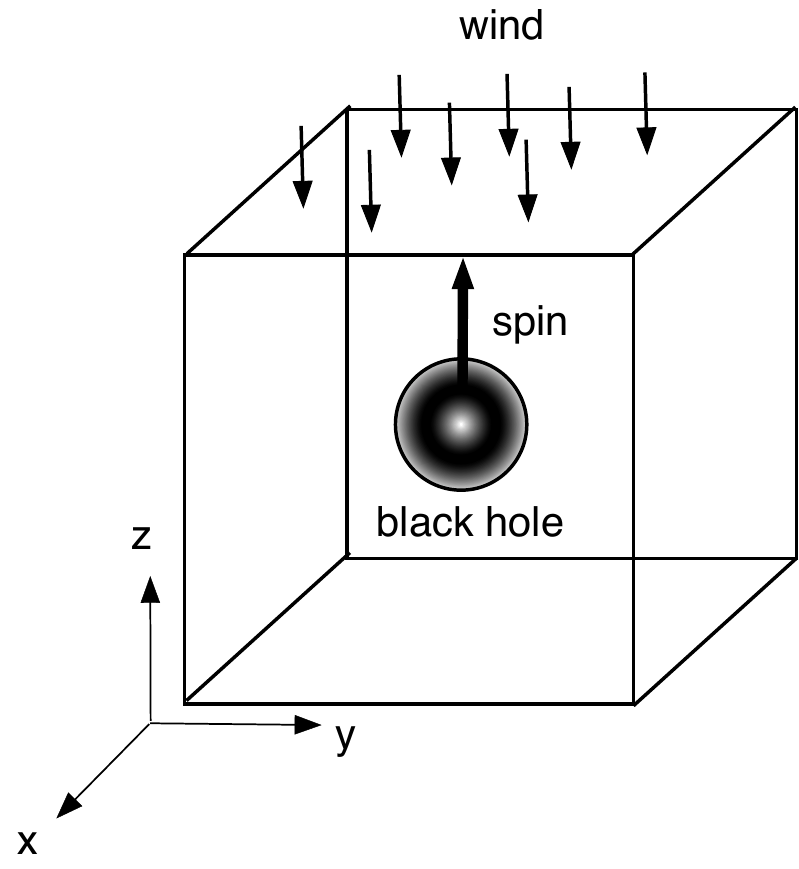}
\caption{Scheme of the accretion process. The cube represents the numerical domain $[-20,20]M\times[-20,20]M\times[-20,20]M$. The wind is injected through the top boundary face with velocity $-v_{\infty}\hat{z}$ and the black hole is at the origin of the domain, with spin parallel to $\hat{z}$. }
\label{fig:cartoon}
\end{figure}

Other specifics of the simulations are that the domain is cubic and we cover it with a fixed mesh refined grid, and that we cover it with three refinement levels with successive halved resolution. We place the black hole at the center of the numerical domain. Instead of moving the black hole across the numerical domain, we inject the wind from the top face of the boundary with velocity $\vec{v}_{\infty}=-v_{\infty}\hat{z}$ toward the black hole, where $v_{\infty}=\sqrt{v_{i} v^{i}}$ is evaluated at the top face. The other five faces of the numerical boundary use outflux boundary conditions and the fluid is allowed to leak through them. The evolution of this process assumes an initially uniform wind density, which we choose to be $\rho_0=10^{-6}$, and with time it evolves toward a stationary regime. The process is illustrated in Fig. \ref{fig:cartoon}. 

In all the cases we assume the speed of sound to be $c_s=0.1c$ and use a wind velocity larger than $c_s$, so that the process is supersonic and eventually forms a stationary shock cone. In order to guarantee that there is a downstream zone we always consider a domain that  contains at least one  sphere of radius $r_{acc}= r_{horizon} / (v_{\infty}^{2} + c_{s}^{2})$, where $r_{horizon}$ is the horizon radius of the black hole, which is the estimated accretion radius in the BHL model \cite{Bondi}. All the simulations are evolved until a stationary regime is achieved and the shock cone remains nearly time independent. In most cases, this is achieved in a time range between 500 and 3000$M$, depending on the asymptotic velocity of the system. The slower the wind, the longer it takes the shock cone to stabilize. 

{\it Velocity regime.} The relative velocity of the black hole with respect to the gas is one of the two parameters the machine learning method has to estimate. In order to  optimize resources, we selected a range of high asymptotic wind speeds. This fact has two important implications: on the one hand, the accretion radius $r_{acc}$ is small and the numerical domain may be small as well; on the other, the shock cone stabilizes faster according to previous experience \cite{Rezzolla,GraciaGuzman}. In this sense we are sacrificing a realistic black hole velocity, or equivalently an asymptotic wind speed  of the order of (for instance) $v_{\infty} \sim 10^3$ km/s ($\sim 10^{-2}c$) as expected for the QSO-3C-186  wandering black hole \cite{QSO3C}. Instead we use a higher speed range of the order $v_{\infty} \sim 10^{-1}c$. Specifically, the range we use to select the simulations is $v_{\infty} \in [0.2,0.8]c$.

{\it Equation of state.} The other parameter we want to estimate from the images  is the adiabatic index of the equation of state, which eventually might help us model the properties of the gas. The range we choose for this parameter is $\Gamma \in [1.1,5/3]$. 

\begin{figure}
\centering
\includegraphics[width= 4.25cm]{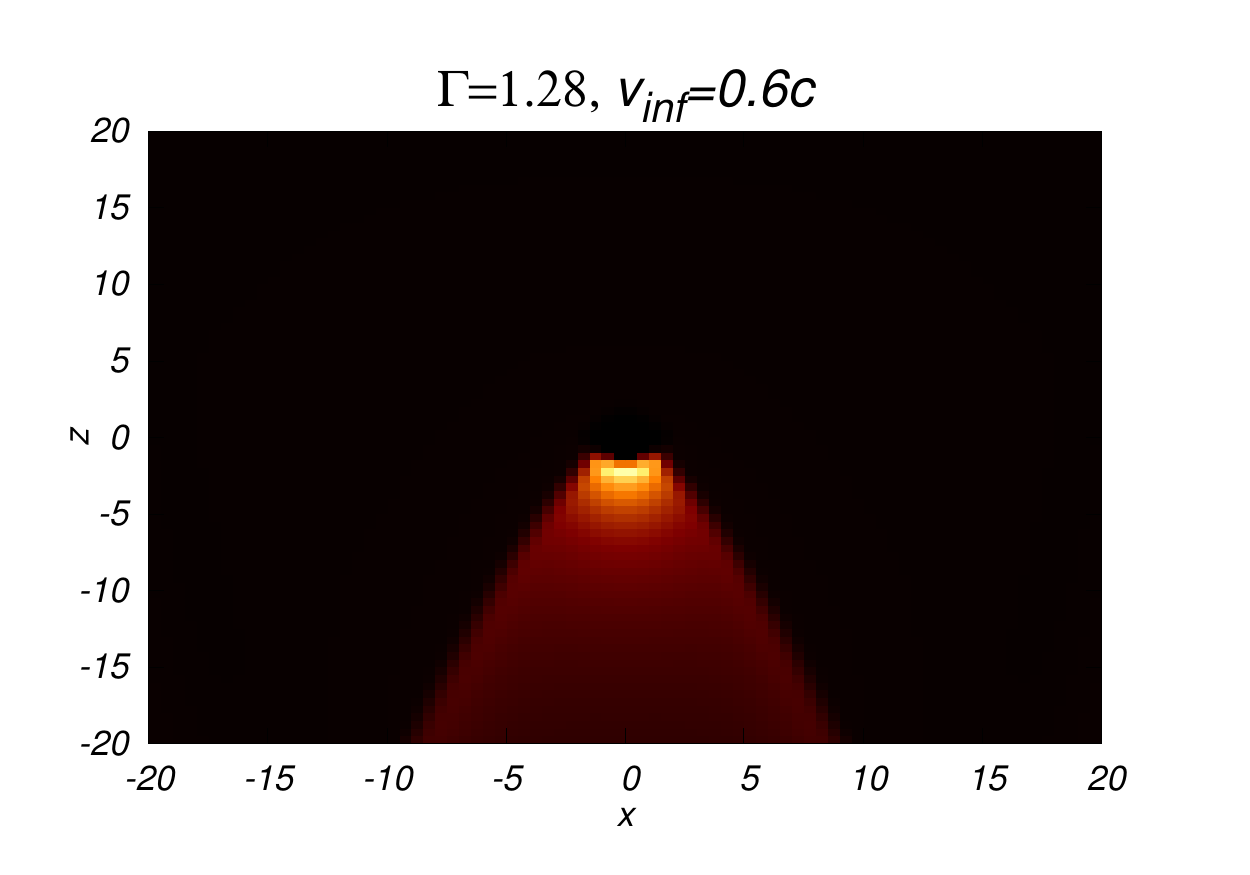}
\includegraphics[width= 4.25cm]{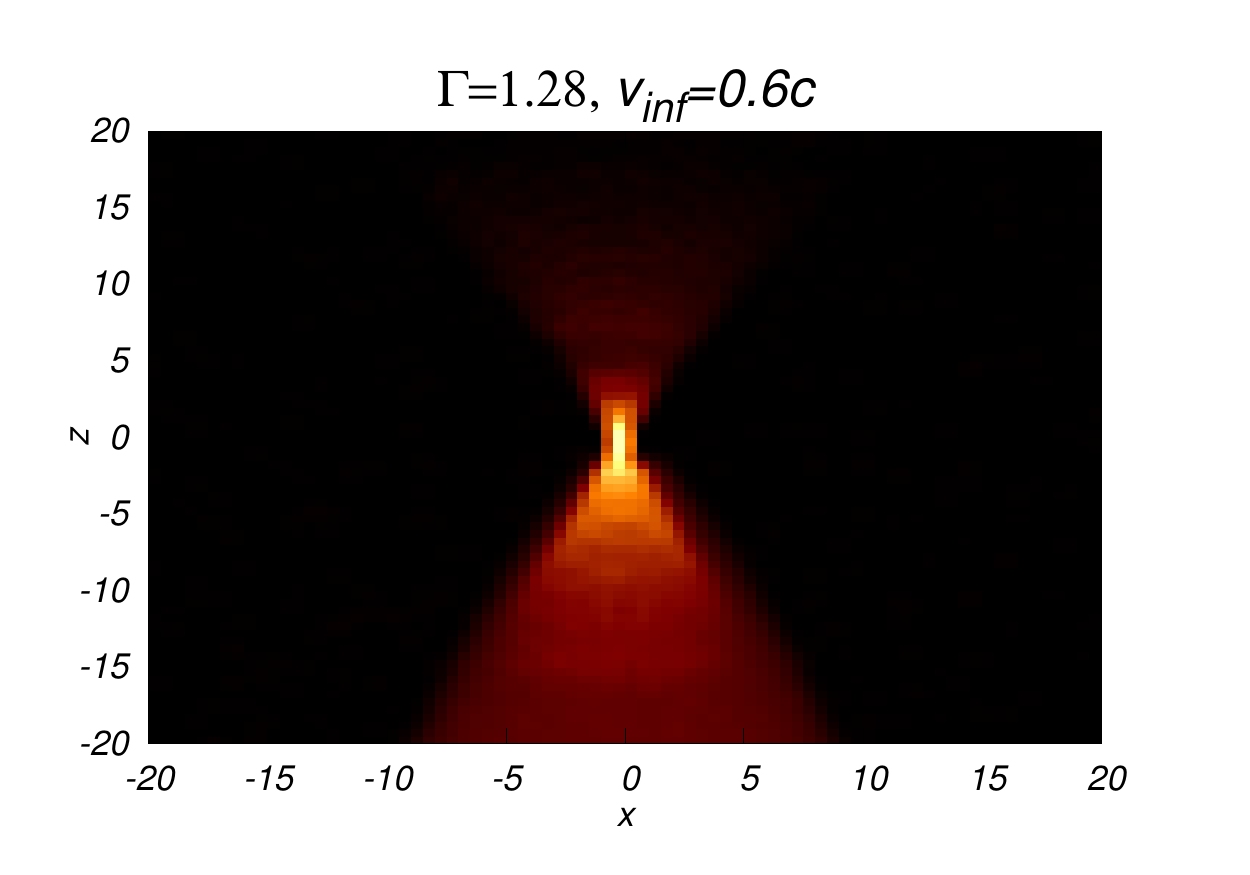}
\includegraphics[width= 4.25cm]{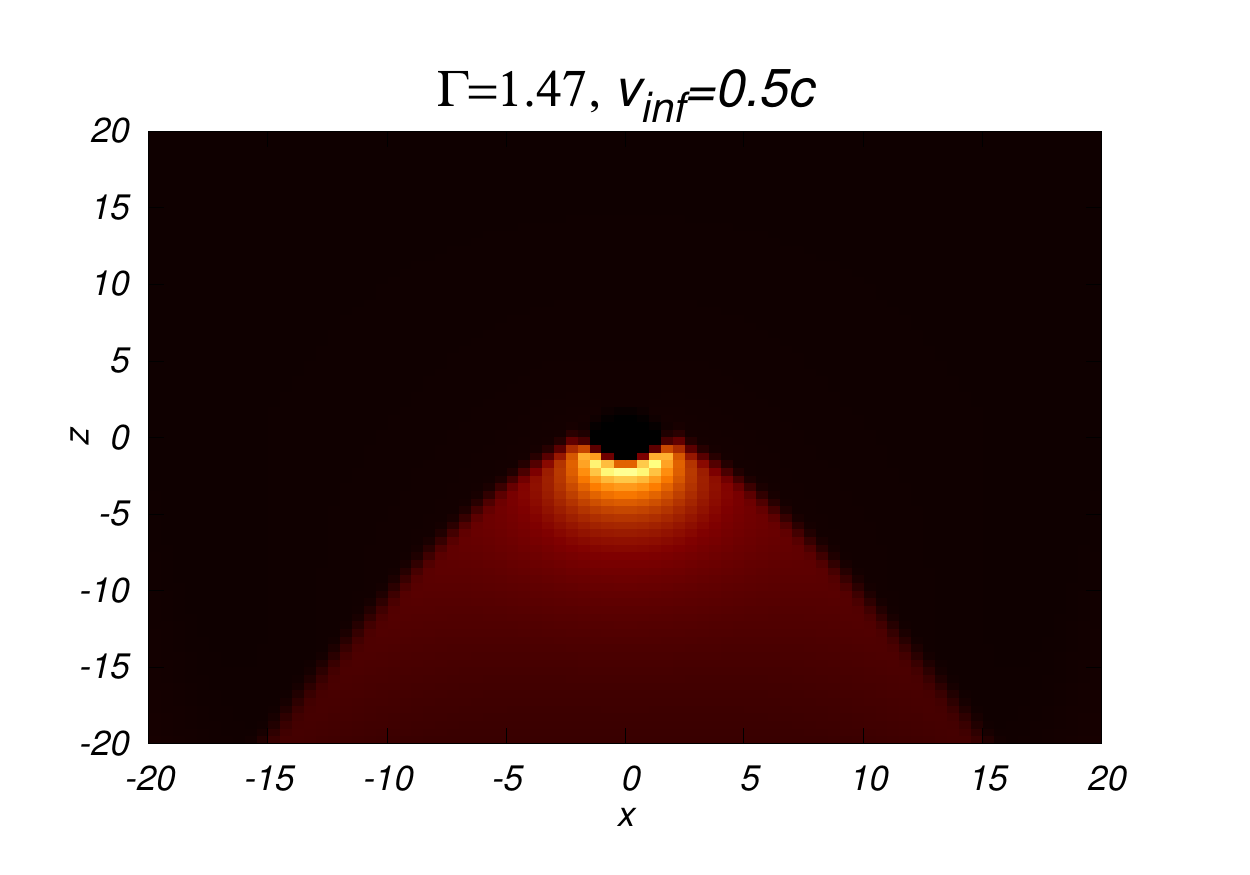}
\includegraphics[width= 4.25cm]{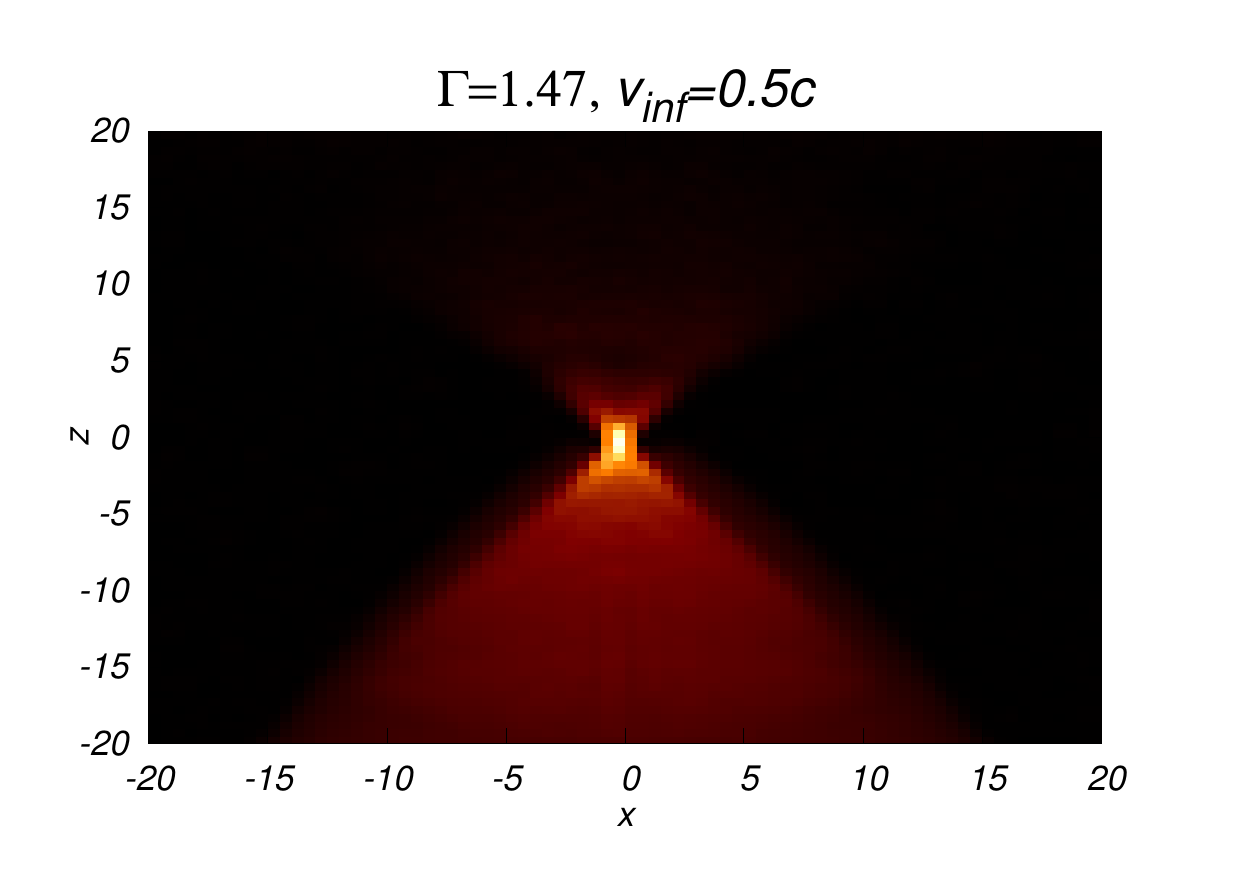}
\caption{Left: We show the typical rest mass density spatial distribution of the high density cone on the $xz$ plane, for two  combinations of the adiabatic index and wind velocity, at a stage in which the process has been stabilized. Right: We show the corresponding lensed images of the density constructed by ray tracing.}
\label{fig:sample}
\end{figure}

We choose these two parameters for our analysis because the properties of the shock cone already show  slightly different profiles depending on the two parameters,  $v_{\infty}$ and $\Gamma$, whereas the influence of other parameters, for instance the spin of the black hole, is smaller within the velocity regime analyzed.  In Fig. \ref{fig:sample} we show the rest mass density on the $xz$ plane for two combinations of $\Gamma$ and $v_{\infty}$, after the density has achieved a stationary regime. In order to capture the strong lensing effects of the black hole on the gas, we generate an image as seen by an observer at 120$M$ with a screen parallel to the $z$ axis. These images were constructed using a ray tracing method based on the solution of null geodesics launched from each volume cell of the gas, and they are shown in the right column of Fig. \ref{fig:sample}. The lensed images in Fig. \ref{fig:sample} are the type of data with which we will train and test the machine learning method. Specifically, the screen of the observer has a resolution of 87$\times$87 pixels. We chose this image resolution in part because it corresponds to the coarsest resolution we use in the simulations, but mainly because the resolution is poor enough  to challenge the pattern recognition capacity of the machine learning method and could potentially be more similar to the resolution of observations.

The parameter space is covered with 900 combinations of a uniformly distributed set of the two parameters in the range $\Gamma \in [1.1,5/3]$ $\times$  $v_{\infty} \in [0.2,0.8]c$. Both dimensions of the domain are split into 30 equally separated values. As can be seen in Fig. \ref{fig:sample}, a method that is capable of recognizing patterns is best able to associate the parameters with which these figures where generated with and the image itself. This is precisely where deep learning methods are useful at. Specifically we use Convolutional Neural Networks (CNNs) that are efficient versions of Artificial Neural Networks (ANNs) used to solve inverse problems in astrophysical scenarios, for example \cite{grupo1,grupo2,grupo3} using simple ANNs and \cite{huerta1,huerta2} using CNNs.

We use the open source code TensorFlow \cite{TensorFlow} running in two GPUs to execute two  CNNs, each of which classifies one of the two physical parameters of the problem. The basic idea behind the classification scheme consists in introducing the values of the rest mass density of each of the 87$\times$87 pixels as the inputs of each CNN. The networks are trained to produce the appropriate outputs that can be interpreted as the possible values for each parameter. The structure of the networks consists of four convolutional layers with ReLu activation functions and each one of them followed by a max-pooling layer. After these eight layers there is a fully hidden connected layer plugged into the final classification layer using a softmax activation function, where both networks use 30 classes as output.
We compute the error of the classification using a cross entropy cost function and minimize this cost function using the stochastic gradient-based optimization algorithm  {\it Adam} \cite{Adam}, with learning rate equal to $0.001$.

\begin{figure}
\centering
\includegraphics[width= 4cm]{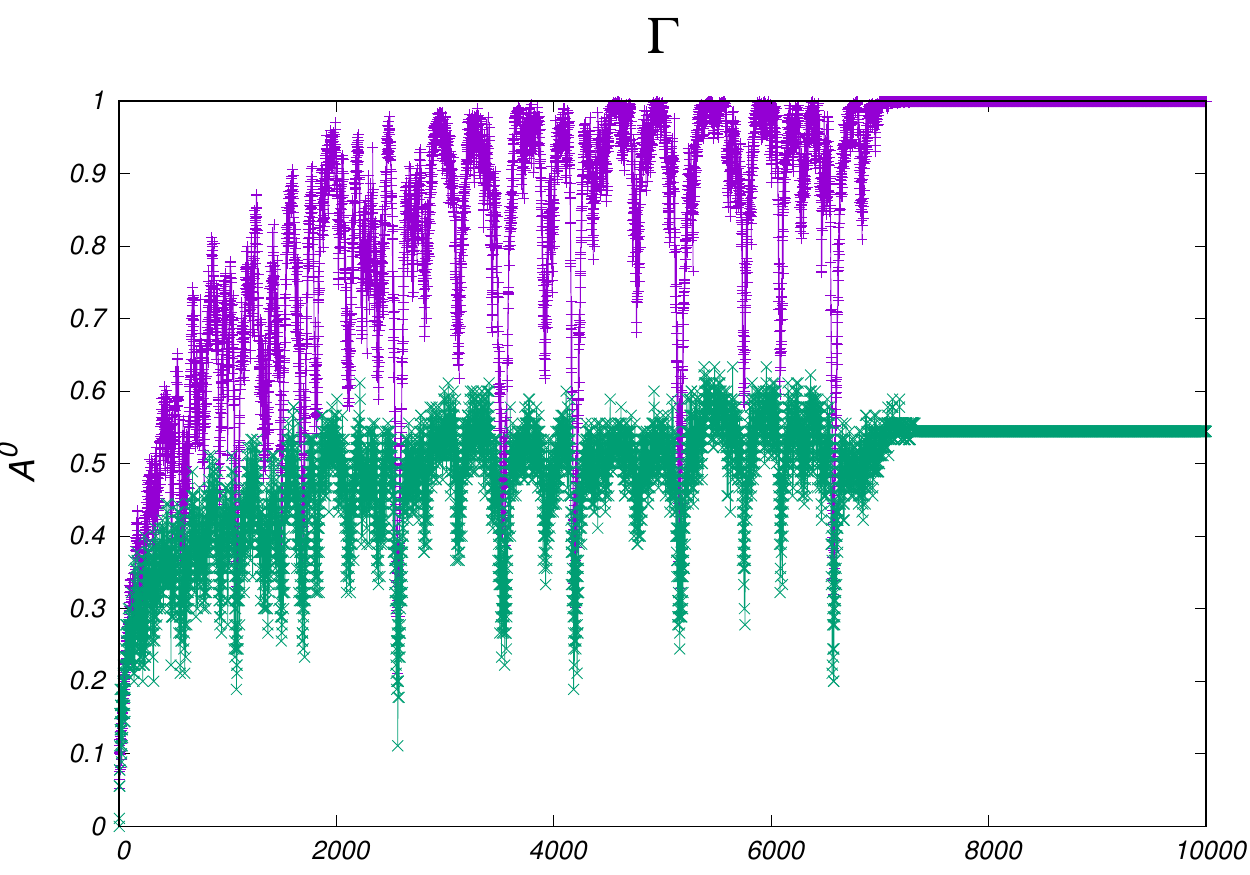}
\includegraphics[width= 4cm]{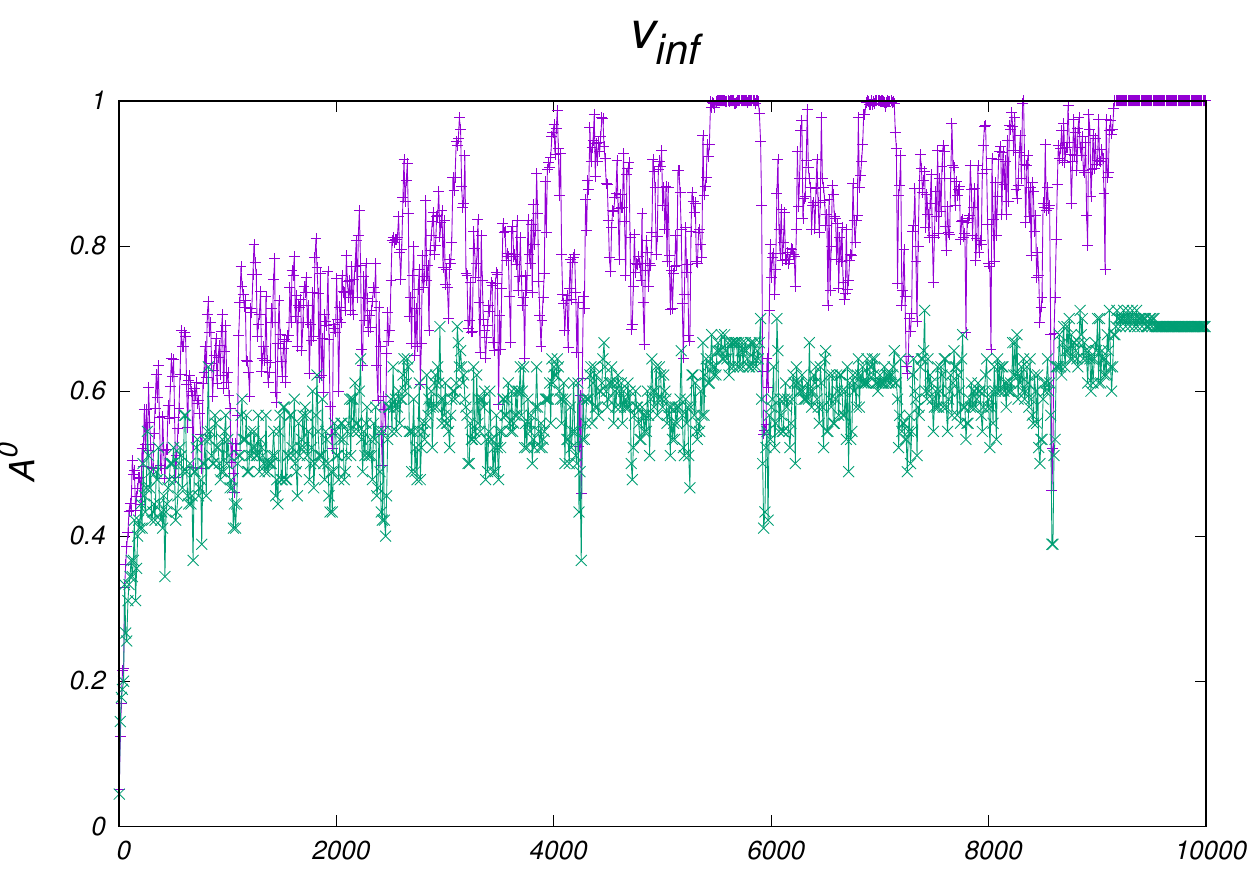}
\includegraphics[width= 4cm]{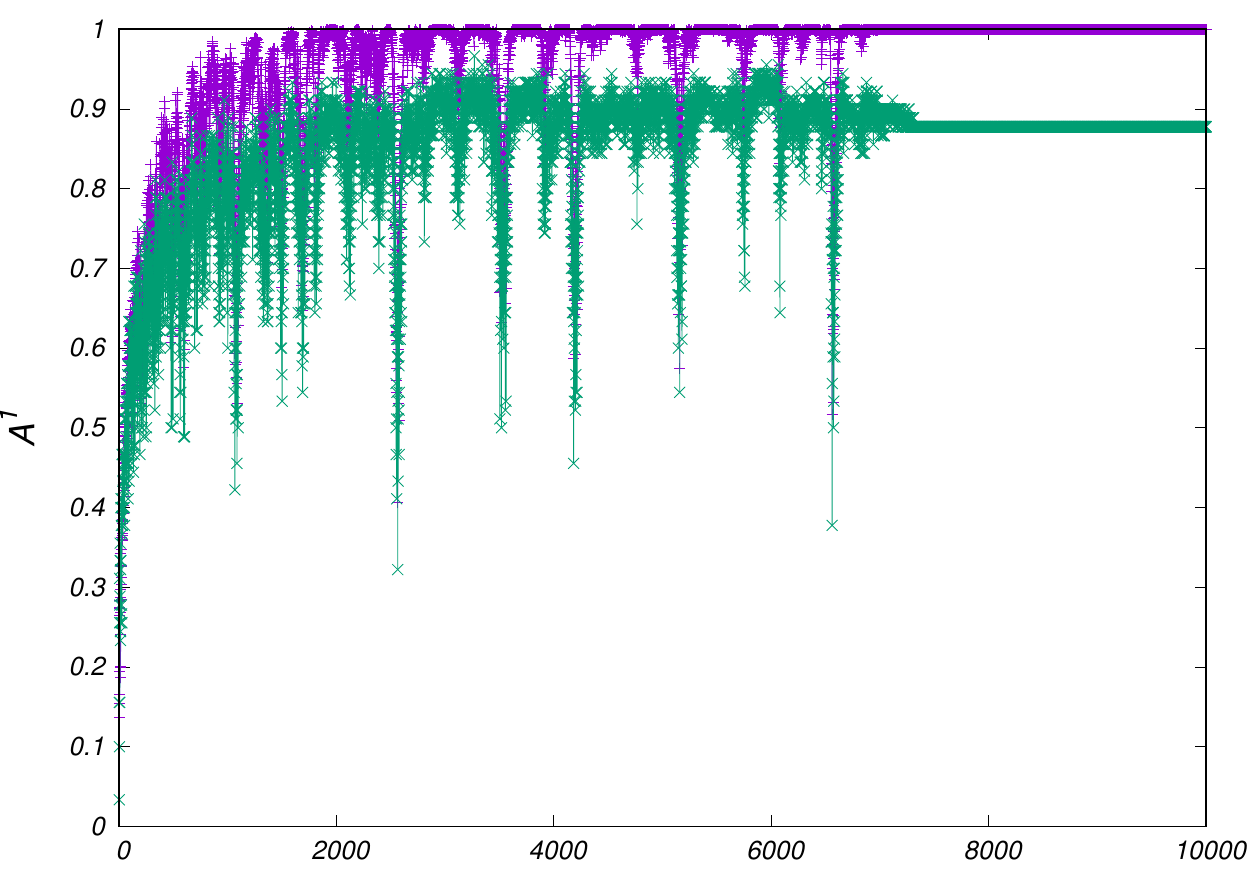}
\includegraphics[width= 4cm]{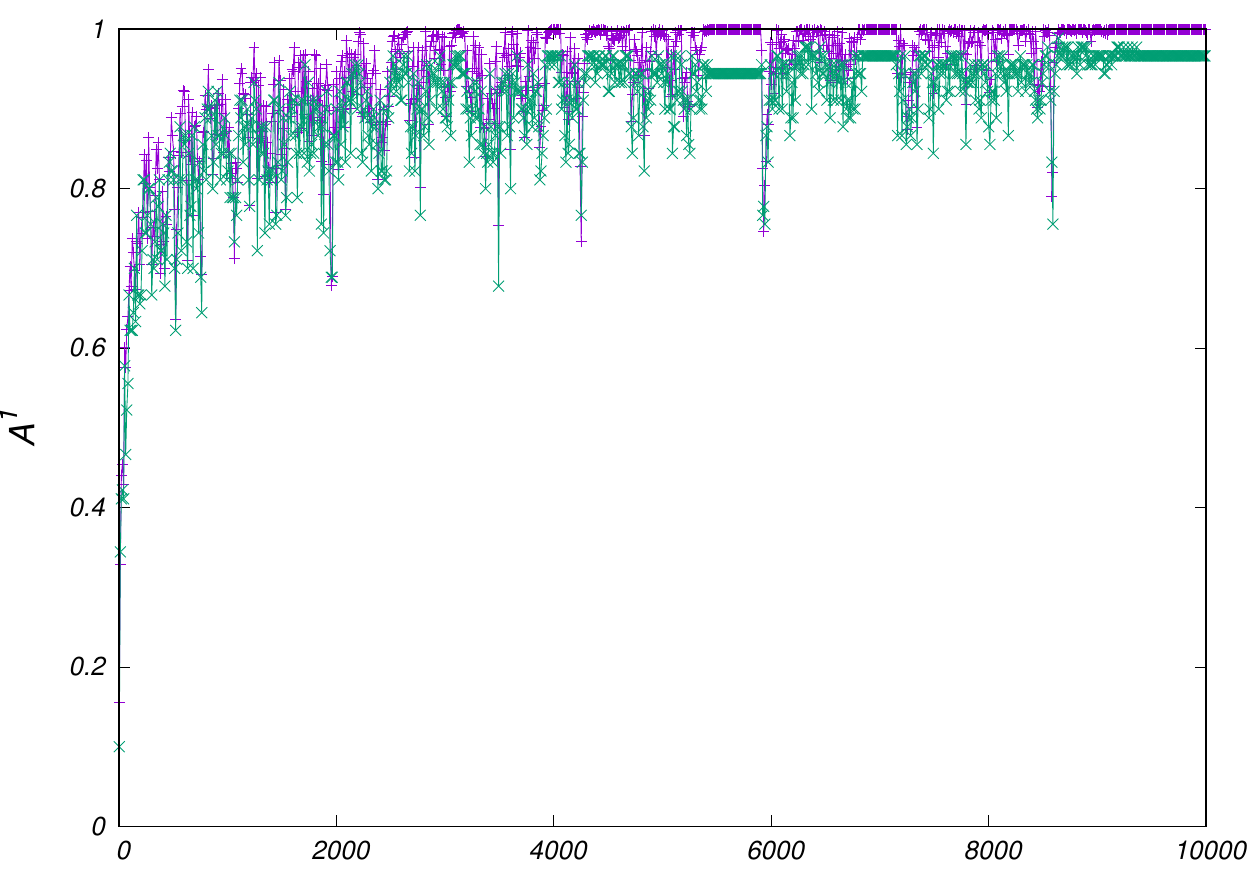}
\includegraphics[width= 4cm]{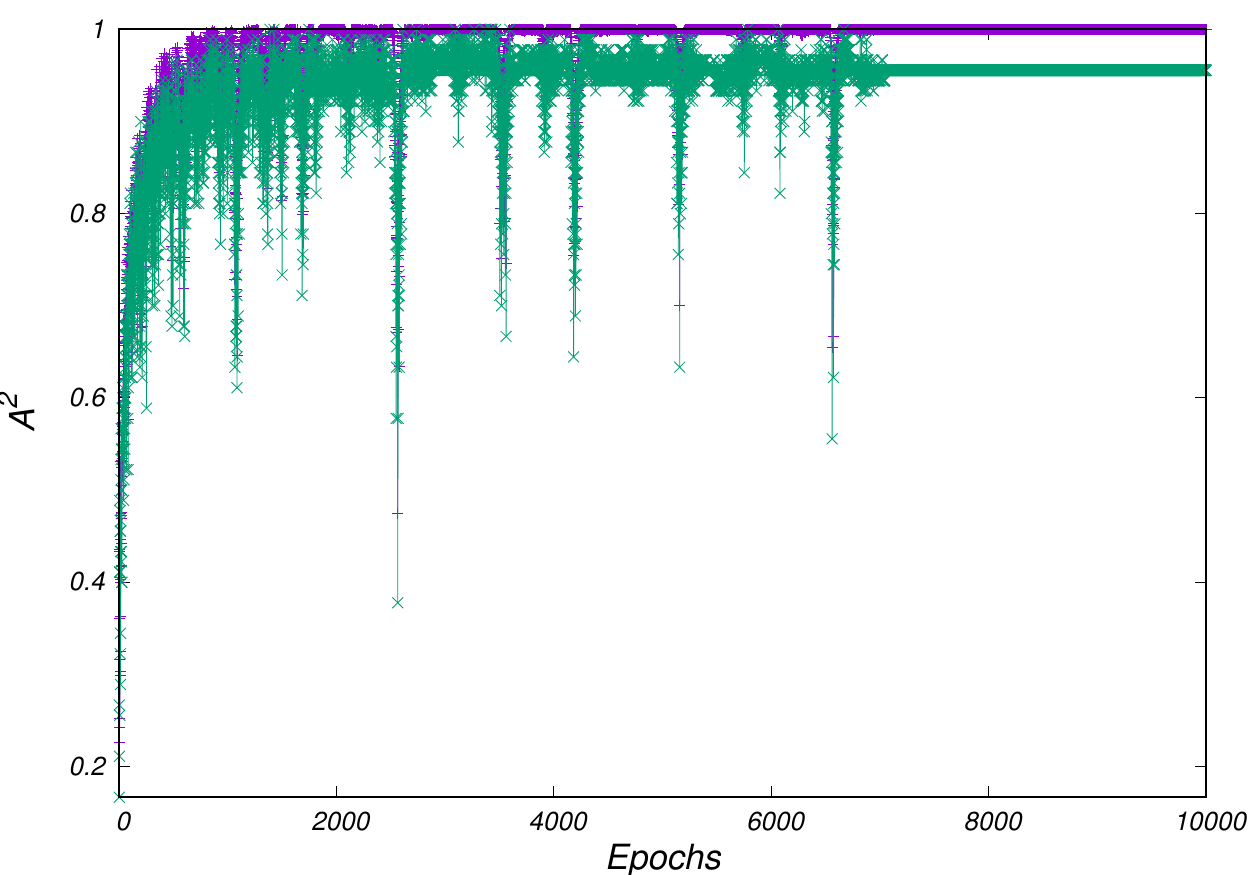}
\includegraphics[width= 4cm]{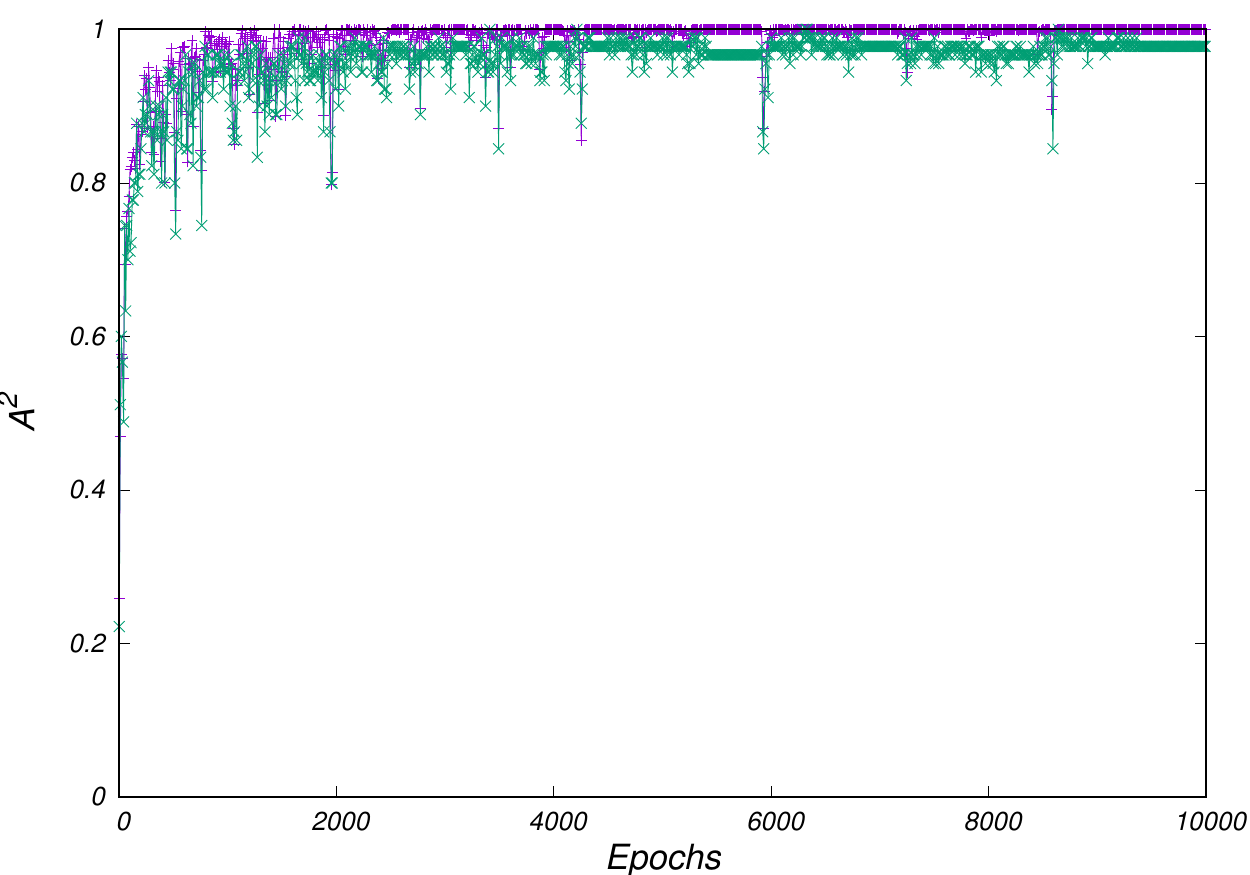}
\caption{Accuracy of the CNN at training and prediction as a function of the number of epochs. Left: $A^0_{\Gamma}, A^1_{\Gamma}$ and $A^2_{\Gamma}$. Right: $A^0_v, A^1_v$  and $A^2_v$. The purple and green lines represent the accuracy of the training and prediction sets.}
\label{fig:Deltas-x}
\end{figure}

Table \ref{table:cnn} presents the parameters describing the best working configuration of the networks.

\begin{table}[htbp]
\begin{center}
\begin{tabular}{|c|c|c|c|c|}
\hline
Name & input & $\#$ filters & output  \\
\hline
Conv 1 & $[1,87\times87]$& 16  & $[16,87\times87]$ \\
\hline
Pool 1 & $[16,87\times87]$& - & $[16,44\times44]$ \\
\hline
Conv 2 & $[16,44\times44]$& 32  & $[32,44\times44]$ \\
\hline
Pool 2 & $[32,44\times44]$& - & $[32,22\times22]$ \\
\hline
Conv 3 & $[32,22\times22]$& 64  & $[64,22\times22]$ \\
\hline
Pool 3 & $[64,22\times22]$& - & $[64,11\times11]$ \\
\hline
Conv 4 & $[64,11\times11]$& 128  & $[128,11\times11]$ \\
\hline
Pool 4 & $[128,11\times11]$& - & $[128,6\times6]$ \\
\hline
FC & $[128,6\times6]$& - & $[1024]$ \\
\hline
Out & $[1024]$& - & $[30]$ \\
\hline
\end{tabular}
\caption{All the convolutional filters have size of $5\times5$ and the pooling layers of $2\times2$. Stride in both cases is equal to one and padding is applied in order to guarantee that the size of the output is equal (half) to the size of the input for the convolutional (pooling) layers. The  final output corresponds to the total number of classes which is equal to $30$.}
\label{table:cnn}
\end{center}
\end{table}

The images of the 900 simulations were divided into two sets: the first one, the {\it training} set, contains 810 simulations chosen randomly from the whole set. Those are the images used to train the network. The remaining 90 simulations correspond to the {\it prediction} set and are used to determine the accuracy of the predictions about data not related to the training. As stated before, we use two different CNNs to perform the classification in each direction. The two networks have the same structure but the learning phase is different in order to classify the corresponding physical parameter. The output of the networks corresponds to the prediction of values of $\Gamma$ and $v_{\infty}$. 

In order to associate a physical value of these parameters and an uncertainty with the selected class, we make the values of the adiabatic index  $\Gamma=1.1094+i \Delta_{\Gamma}$ with $\Delta_{\Gamma}=0.0189$, and we make the values of the velocity  $v_{\infty}=0.21c+i\Delta_{v}$ with $\Delta_{v}=0.02c$ and $i=0,...,29$. 

We present the complete training and prediction sets to the network and count $N_0$ the number of classifications that correspond exactly to the desired class. Then we count $N_{-1}$($N_1$), the number that missed the correct classification by one class to the left (right).  We continue doing this up to $N_{-29}$($N_{29})$.
Dividing each of these values by the total number of elements of the set ($N_{set}=810$ for training and $N_{set}=90$ for prediction), we compute the percentage of elements of the set that were classified correctly within an interval of $n$ neighboring classes centered on the correct one. We call this number $A^n$ and it is explicitly computed as 

\begin{equation}
A_{\Gamma, v}^n = \frac{1}{N_{set}} \sum^n_{i=-n} N_i . \nonumber
\end{equation}

\noindent The corresponding results for $A^{0,1,2}_{\Gamma}$ and $A^{0,1,2}_v$ are presented in Fig. \ref{fig:Deltas-x}  in the left and right columns correspondingly. 
A quick analysis shows that the parameter $\Gamma$ is much easier to learn for the CNN compared with $v_{\infty}$. The number of learning epochs to obtain comparable results for both parameters is six times bigger for the velocity than for the adiabatic index.

{\it Prediction estimates.}  Each of the two classes predicted by the CNNs selects an interval for each physical parameter. For instance, if class 1 is selected for parameter $v_{\infty}$ the values contained in that class range from $0.2c$ to $0.2c+\Delta_v=0.22c$. So, we interpret the prediction of the networks as the mid-value of the interval, in this case the prediction would be $ v_{\infty}=0.21c$. In order to settle the uncertainty and the confidence interval on the prediction, we use the length of the interval ($\Delta$) and the quantities $A^n$. On the one hand, looking at the left-hand plots in Fig. \ref{fig:Deltas-x}, we can say that the prediction of the CNN for the parameter $\Gamma$ is correct $54.44\%$ of the time, with an uncertainty of $\pm 0.0095$; $87.78\%$ of the time, with an uncertainty of $\pm 0.0284$; and $95.56\%$ of the time, with an uncertainty of $\pm 0.0473$. On the other hand, looking at the right-hand plots in Fig. \ref{fig:Deltas-x}, we can conclude that the predicted value for the parameter $v_{\infty}$ is correct $68.89\%$ of the time, within an uncertainty of $\pm 0.01c$; $96.67\%$ of the time, with an uncertainty of $\pm 0.03c$; and $97.78\%$ of the time, with an uncertainty of $\pm 0.05c$. 

So far, we have presented a method that classifies with high accuracy two important parameters of a wandering black hole system. In this first study we use images of the rest mass density that are ray-traced into a screen assuming the gas itself has no opacity. This method suffices to take the strong lensing into account and in a future more sophisticated version, it could help us create a more elaborate model  of matter. We should also implement sophistications based on a catalog constructed with the simulations using velocities in a range of more realistic scenarios such as, potentially the candidate QSO 3C 186 \cite{QSO3C} \cite{Lousto2017}. 

Our method and type of catalog open the door to studying other parameters involved in  accretion processes on black hole space-times and we expect it to be efficient enough for the analysis of the EHT observations. Our analysis and tools are also expected to be important in the study of the premerger phase of binary supermassive black holes, where the dynamics is mostly driven by the interaction of  individual black holes with matter, before the interaction between the holes becomes general relativistic  \cite{Dosopoulou2017}. In fact, these conditions are currently being used to estimate the Gravitational Wave background sourced by these binaries, which highlights the current interest of accretion processes on moving black holes \cite{Kelly2017}.

In order to explore how applicable our method could be in scenarios involving parameters similar to those of QSO 3C 186, we performed two simulations that help one to have a picture of the properties of the system and the difficulties that must be sorted out to construct a catalog for this astronomical object. For this we assume the mass of the black hole is  $3\times 10^9 M_{\odot}$, with a velocity of 2100km/s  and dimensionless spin parameter of $0.91$ as modeled in \cite{Lousto2017}. We devise two different scenarios in the supersonic and subsonic regimes.

\begin{figure}
\includegraphics[width= 4.25cm]{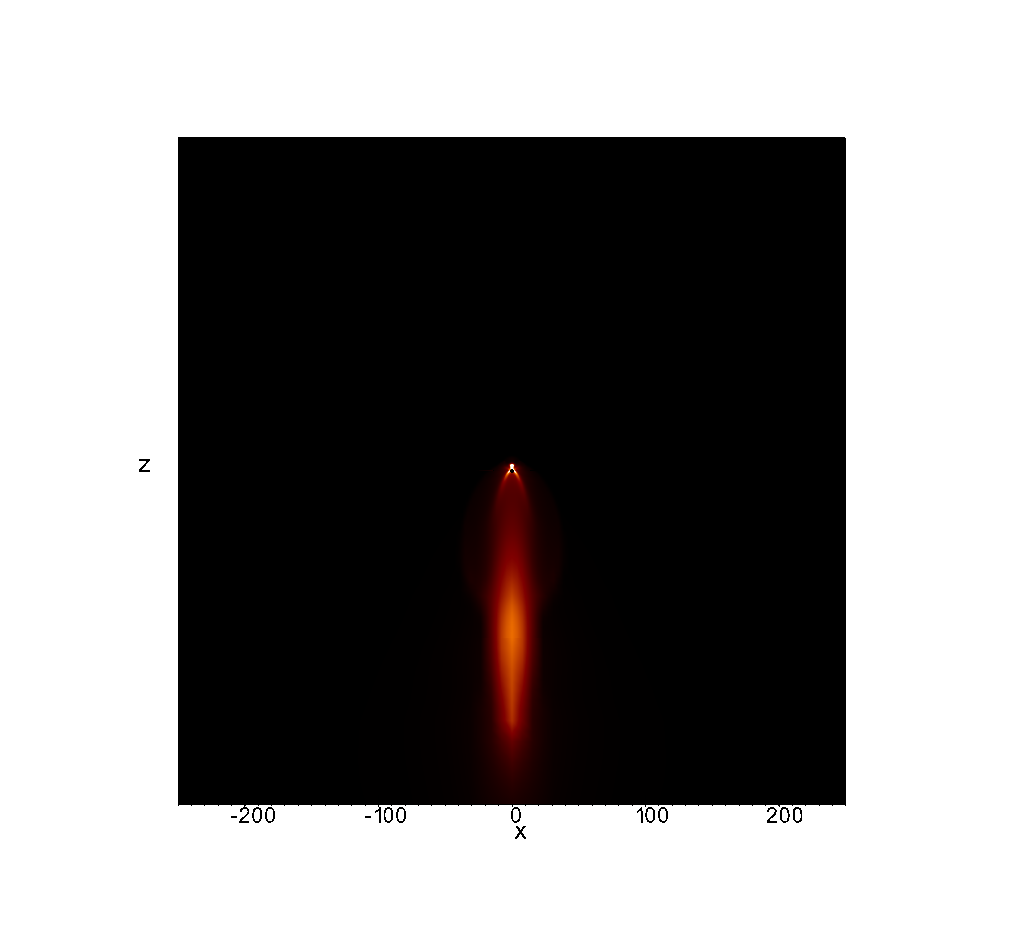}
\includegraphics[width= 4.25cm]{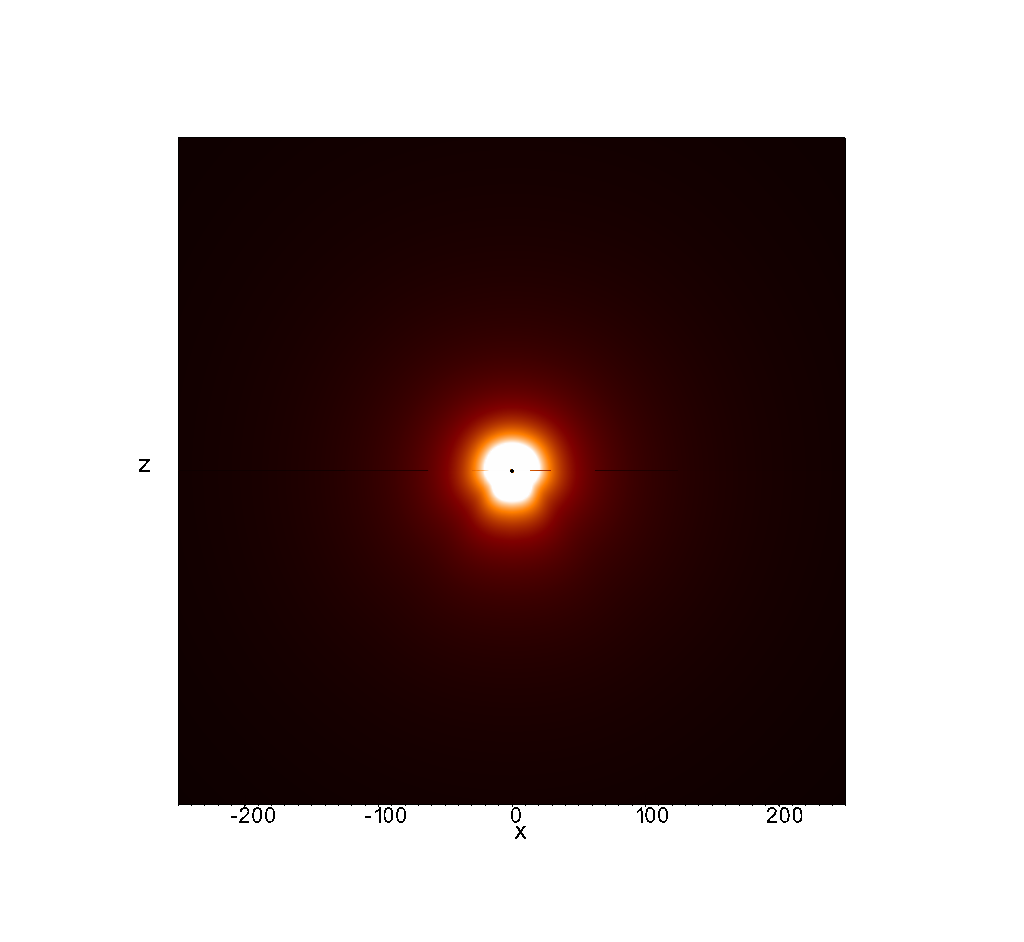}
\caption{We show the rest mass density over the $xz$ plane, for the supersonic and subsonic cases with the parameters of the QSO 3C 186 candidate. The black hole can barely be seen and the lensing effects are expected to be smaller in this scenario.}
\label{fig:qs0136}
\end{figure}

Supersonic scenario. We assume the speed of sound to be $c_{s,\infty}=0.0035 {\rm c}$. The asymptotic wind velocity is set to $v_{\infty}=2100km/s \sim 0.007 {\rm c}$, so that the Mach number is ${\cal M}_{\#}=2$. For this case we assume $\Gamma=4/3$. The accretion radius in units of $M$ is therefore $r_{acc} = 1/(c_{s,\infty}^2 + v_{\infty}^2) \sim 16327M$.  In order to contain a sphere of this radius we set the numerical domain to be $[-20000M,20000M]^3$, which  guarantees that there are upstream and downstream regions. We covered the numerical domain using 13 refinement levels, each one being a cube nested at the center of the other. The resolution of the highest level is 0.1526M covering a box of size $[-4.6M,4.6M]^3$, which helps us capture the details of the accretion process near the black hole's horizon. The time it takes the wind to relax into a nearly time independent configuration is expected to be of the order of a crossing time of the system; for this we use the speed of sound (which is half that of the wind) and estimate the crossing time to be $t_{ct} \sim 1.14\times 10^6 M$. This long evolution time together with the number of nested refinement levels makes this simulation very costly, which is the reason why the construction of a catalog of hundreds of simulations is not currently available. The resulting distribution of rest mass density after the system becomes stationary is shown in Fig. \ref{fig:qs0136}.

Subsonic scenario. We assume the speed of sound to be $c_{s,\infty}=0.028 {\rm c}$ so that the asymptotic Mach number is ${\cal M}_{\#}=0.25$. For this case we assume $\Gamma=5/3$. The accretion radius in this case is $r_{acc} = 1/(c_{s,\infty}^2 + v_{\infty}^2) \sim 1200M$ and we set the numerical domain to be $[-2000M,2000M]^3$. In this case we used 10 refinement levels to cover the domain, with the highest resolution 0.12M covering a box of size $[-4.55M,4.55M]^3$. The crossing time is of order $t_{ct} \sim 1.66\times 10^5 M$, a smaller time compared to the supersonic case. The density distributes differently, the bow shock compensates the upstream flow approaching the black hole and the stationary regime shows a rather spherical distribution as shown in Fig. \ref{fig:qs0136}.

Notice that in these two simulations the black hole can barely be seen and the effects of lensing are expected to be smaller. These scenarios show various features. First they show how expensive they are in terms of computer power, resolution, domain size and  time of evolution. Second, they show how  the distribution of matter changes from one scenario to the other; notice that  in the supersonic case we use $\Gamma=4/3$ and in the second $\Gamma=5/3$, with two different sound speeds, despite the fact that the wind travels at the same velocity of $0.007 c$. Finally, the images show that the differences between the two distributions are significant enough as to be captured by our CNN. Thus, we expect that the method presented in this paper can potentially track down the properties of the gas of systems similar to those of the realistic case QSO 3C 186 once higher resolution images are available.


\section*{Acknowledgments}
This research is supported by Grants No. CIC-UMSNH-4.9 and CIC-UMSNH-4.23,  and by CONACyT Grant No. 258726 (Fondo Sectorial de Investigaci\'on para la Educaci\'on). Part of the simulations were carried out in the computer farm funded by CONACyT 106466 and the Big Mamma cluster in our laboratory.  The authors also acknowledge the computer resources, technical expertise and support provided by the Laboratorio Nacional de Superc\'omputo del Sureste de M\'exico, CONACyT network of national laboratories. We also thank ABACUS Laboratorio de Matem\'aticas Aplicadas y C\'omputo de Alto Rendimiento del CINVESTAV-IPN, Grant No. CONACT-EDOMEX-2011-C01-165873, for providing computer resources.


\end{document}